%% file: PMiddle.tex
\documentclass[preprint,12pt]{elsarticle}

\usepackage{amssymb}

\begin{document}

\begin{frontmatter}


\author{L. F. Johnson}
 \ead{lfjsde@gmail.com }
\pdfoutput=1
\title{Middle and Ripple, fast simple $O(\lg n)$  algorithms for Lucas Numbers}


\address{University of Waterloo}

\begin{abstract}
A fast and simple  $O(\log n)$ iteration algorithm for individual Lucas  numbers 
is given. This is faster than using Fibonacci based methods because of the structure of Lucas numbers. Using a $\sqrt 5$
conversion factor gives a faster Fibonacci algorithm because the speed up proposed in [5] 
also directly applies.

A fast simple   recursive algorithm for individual Lucas  numbers 
is given that is $O(\log n)$. 
 
\end{abstract}


\end{frontmatter}



\input{LucasMiddle.tex}





\bibliographystyle{elsarticle-num}
\bibliography{<your-bib-database>}





\end{document}

%% file: LucasMiddle.tex

\section{Introduction}
When initial conditions for the Fibonacci recurrence, $\ F_{n}= F_{n-1} +F_{n-2}$,  
are changed to 1,3 the result is the Lucas numbers (1891), $\ L_{n}= L_{n-1} +L_{n-2}$,
with $L_{1}=1$, $L_{2}=3$. It is convenient to define $L_{0}=2$, 
then $\ L_{2}= L_{1} +L_{0}$.  

There are a number of iterative solutions for this 
recursive definition, using addition,  that are $O(n)$ [6], the lower bound for enumerating 
the sequence; however, individual numbers can be found in sublinear time.
DeMoivre published a closed formula in 1730 that requires $n$ multiplications [1].    
$F_{n}$ was first shown to be found in $O(\log n)$ time in 1978 [3 ] followed by 
improved algorithms [4,5]. None of these dealt directly with Lucas 
numbers perhaps because  $\ L_{n}= F_{n-1} +F_{n+1}$, requiring two calls. 

For simplicity, we need not discuss  
the size of $L_{n}$, which  grows as $O(F_{n})$, because this is the same for all 
algorithms (when the bit model executions costs are excluded [4,5]). Execution costs are 
based on the represention operations, such as number of additions and multiplications.


The basic idea for obtaining an $O(\log n)$ algorithm  is the use of doubling formulas.
Simple doubling only gives values of $n$ that are powers of $2$. 
Some manipulation is required to to solve for all $n$.
Using the well known identities [1], $ F_{n+m} = F_{m} * F_{n+1}+ F_{m-1}*F_{n}$ 
and   $ F_{n+1} * F_{n-1} =  F_{n}^2 + (-1)^n$ , algorithms can be developed for $F_n$. 
A clever approach is to combine the related Fibonacci and Lucas sequences as 
for example, $\ F_{2n}= F_{n} * L_{n}$ and $\ L_{2n}= L_{n}^2 - 2$, where $2n$ is even [3,4,5].
A single formula can only be used to find $L_{n}$ when $n$ is a power of 2 and 
some manipulation is required to obtain a general solution for all $n$.

Noting that the general doubling formula for Lucas is actually $\ L_{2k}= L_{k}^2 - {(- 1)}^k*2$
and substituting k+1 for k  
then gives $\ L_{2k +2}= L_{k+1}^2 + {(- 1)}^k*2$. This single fundamental equation form is then 
the basis for a compact solution compared to [5].

\section{ Iteration}
To calculate $L_{n}$ for any $n$ requires three adjacent terms in the sequence.
Using a method based on a recursive based calculation the two terms to double are 
selected by markOdd() in the following algorithm.

\subsection{Middle}
For simplicity, the sequence variables are renamed as follows: 
 $LL = L_{2k}, LM = L_{2k+1},   LH = L_{2k+2}  $
The following formulas are used to double until $L_{n}$ is found. 

LL= LL*LL - p*2 \\
LH= LM*LM + p*2 \\
LM= LH-LL \\

Depending on the step, two different updates are required to prepare for the next step 
these are selected using an odd/even tag. 
This can be seen in the algorithm. 

Two square multiplications, two sign calculations, and three additions in the iterative loop
 make this very fast compared to [5]. 
No pre conditions nor post conditions are used as was the case in [5]. All general methods need some kind 
of odd(i) selected calculation, when n is not a power of 2, and this has been reduced 
here to the insertion of a shift forward of two sequence terms and one assignment.

The algorithm can be expressed as a recursion in a direct way, which is done in the next section.
Using the relation, $F_n = \lceil( L_n/ \sqrt 5 -0.5)  $ , our algorithm also  gives a very fast and simple 
calculation of a Fibonacci number.

$$
\begin{array}{l}
 Algorithm: \ Middle(n) \ given \ n > 1, \ return \ LL=L_{n}   \\
     N \leftarrow \lfloor \lg n  \\
    
     array  \ markOdd[N] \leftarrow 0                          \\
     i \leftarrow  n ;  j \leftarrow N                                    \\ 
   while(j >0)\{if ( odd(i)) \ markOdd(j)= 1; i= i/2;j=j-1 \} \ \\
       
     j  \leftarrow 1\\                     
     while (j\leq N-1) \ \  \\
   \ \ \ \     p \leftarrow 1              \\
 \ \ \ \    if ( markOdd(j)) \{LL \leftarrow LM; LM \leftarrow LH;
                                 p \leftarrow -1 \  \} \\
     \ \ \ \      LL \leftarrow LL * LL -p*2           \\
    \ \ \ \      LH  \leftarrow LM*LM+ p*2    \\
    \ \ \ \      LM \leftarrow  LH-LL          \\
     
      \ \ \ \  j \leftarrow j+1; \ \ \	\\ 
    endwhile \\
    if ( markOdd(j)) \ LL \leftarrow LM    \\

    return \ LL    
\end{array}
$$

 \section{Recursion}
The recursion  $\ L_{2n}= L_{n}^2 - 2$ when n is even 
can  be used to find $L_{n}$ in $O(\log n)$ time when $n$ is a power of 2. 
This can be executed recursively by using $\ L_{k}= L_{k/2}^2 - 2$ 
The general form of this doubling formula for Lucas is  
$\ L_{2k}= L_{k}^2 - {(- 1)}^k*2$ [1] and substituting k+1 for k  
then gives $\ L_{2k +2}= L_{k+1}^2 + {(- 1)}^k*2$. The result on two adjacent position terms 
is two even position semi-adjacent terms.
To obtain $\ L_{n}$  when $n$ is odd, calculate $\ L_{n+1} -\ L_{n-1}$.  
Thus to calculate $L_{n}$ for any $n$ requires three adjacent terms in the sequence.

\subsection{Ripple}

For simplicity, the sequence variables are renamed as follows: 
 $LL = L_{k}, LH = L_{k+2}  $
The following formulas are used to double until $L_{n}$ is found. 

LL= LL*LL - p*2 \\
LH= LM*LM + p*2 \\

Depending on the odd and even values of $n$ and $n/2$,
different updates are required. The correct equation to use is
selected by the variable $p$, as can be seen in the algorithm. 

$$
\begin{array}{l}
 Algorithm: \ Ripple(n) \ given \ n > 1, \\ 
return \ L_{n}   \\         
     if ( n=2) \ return \ 3       \\
     if ( n=3) \ return \ 4       \\ 
     if ( n=4) \ return \ 7       \\
     if \ (even(n/2)) \ then \ \ p \leftarrow \ 1 \ else \ p \leftarrow -1  \\
     LL \
\leftarrow Ripple( \lfloor n/2 \rfloor ) \\
     if \ ( even(n)) \ then \ return \ LL * LL -p*2  \\
       \ \ \ \        else \ LH \leftarrow Ripple( \lceil n/2 \rceil) \\
       \ \ \ \             return \ LH * LH - LL*LL +p*4  \\
      \end{array}
$$

If a compiler is able to remember multiple identical calls, the algoritm can be simplified 
to remove LH and HL.

$if \ ( even(n)) \ then \ return \ L(n/2)^2 -p*2  \\
       \ \ \ \        else \         return \ L(\lceil n/2 \rceil)^2 - 
L(\lfloor n/2 \rfloor )^2  +p*4  $

Each call uses a selection of $p$, one or two square multiplications, and one signed addition. 
 Making  this very fast compared to [5], when the cost of recursion is ignored. 
No pre conditions nor post conditions are used as was the case in [5], except to define the 
recursion basis.

\section{Conclusions}
For linear Fibonacci algorithms, a change in initial conditions gives a Lucas algorithm.
Any $O(\log n)$ Fibonacci algorithm can be used to find the Lucas number indirectly by the 
relation $\ L_{n}= F_{n-1} +F_{n+1}$. While this requires two external calls, these algorithms 
can usually be modified internally to produce  $ L_{n}$ in a single call 
(which may only hide the two external calls). Conversely, 
Fibonacci numbers can be derived from Lucas numbers using a $\sqrt 5$ divison factor.
By the nature of the simple foundation relation of Lucas squares,
our algorithm is simpler and faster than other Fibonacci algorithms. 

Since our algorithm only uses two square multiplication per iteration, the speed argument 
for squares in [5], claimed to be the fastest  
for very large $n$, also applies to our method.
This algorithm is much simpler and faster by a constant factor than that in [5]
(which did not internally implement 
the fast FFT-based Schoenhage-Strassen process for fast multiplication). 
Thus, the two algorithms can be directly compared.

In Ripple the the average number of multiplications is less than 2. 
The algorithm only uses squares, and the argument in [5], claimed to be the fastest  for 
very large $n$, also applies to this method.

While Lucas numbers can be derived from finding Fibonacci numbers, it is always more efficient 
to calculate the Lucas numbers directly.